\documentclass[10pt]{article}
\usepackage{graphicx}
\begin{document}

\begin{center}
{\Large \bf Wigner Rotations and Iwasawa Decompositions in
Polarization Optics}

\vspace{7mm}

D. Han\footnote{electronic address: han@trmm.gsfc.nasa.gov}
National Aeronautics and Space Administration, Goddard Space
Flight Center,\\ Code 935, Greenbelt, Maryland 20771 \\

\vspace{5mm}

Y. S. Kim\footnote{electronic address: yskim@physics.umd.edu}\\
Department of Physics, University of Maryland,\\
College Park, Maryland 20742, U.S.A.\\

\vspace{5mm}

Marilyn E. Noz \footnote{electronic address: noz@nucmed.med.nyu.edu}\\
Department of Radiology, New York University, New York,
New York 10016, U.S.A.

\end{center}

\begin{abstract}
Wigner rotations and Iwasawa decompositions are manifestations of
the internal space-time symmetries of massive and massless particles,
respectively.  It is shown to be possible to produce combinations of
optical filters which exhibit transformations corresponding to Wigner
rotations and Iwasawa decompositions.  This is possible because the
combined effects of rotation, phase-shift, and attenuation filters
lead to transformation matrices of the six-parameter Lorentz group
applicable to Jones vectors and Stokes parameters for polarized
light waves.  The symmetry transformations in special relativity
lead to a set of experiments which can be performed in optics
laboratories.

\end{abstract}

\section{Introduction}
In our earlier papers~\cite{hkn97josa,hkn97pre}, we have formulated
the Jones vectors and Stokes parameters in terms of the two-by-two
and four-by-four matrix representations of the six-parameter Lorentz
group~\cite{brown95}.  It was seen there that, to every two-by-two
transformation matrix for the Jones vector, there is a corresponding
four-by-four matrix for the Stokes parameters.  It was found also that
the Stokes parameters are
like the components of Minkowskian four-vectors, and two-component
Jones vectors are like two-component spinors in the relativistic world.
This enhances our capacity to approach polarization optics in terms of
the kinematics of special relativity.

Indeed, we can now design specific experiments which will test some
of the consequences derivable from the principles of special relativity.
The most widely known example is the Wigner rotation.  This has been
extensively discussed in the literature in connection with the
Thomas effect~\cite{hks87cqg}, Berry's phase~\cite{chiao88,kitano89},
and squeezed states of light~\cite{knp91}.

In our earlier papers, we discussed an optical filter which will
exhibit the matrix form of
\begin{equation}\label{shear1}
\pmatrix{1 & u \cr 0 & 1}
\end{equation}
applicable to two transverse components of the light wave, where u is a
controllable parameter.  When applied to a two-component system, this
matrix performs a superposition in the upper channel while leaving the
low channel invariant.  The question is whether it is possible to
produce optical filters with this property.

In Ref.~\cite{hkn97josa}, we approached this problem in terms of the
generators of the Lorentz group.  It is very difficult, if not
impossible, to manufacture optical devices performing the function of
group generators.  In the case of optical filters, this means an
infinite number of layers of zero thickness.  In the present paper,
we deal with the same problem from an experimental point of view.
We will present a specific design for optical filters performing this
function.  We will of course present our case in terms of a combination
of three filters of finite thickness.

In order to achieve this goal, we use the fact that polarization optics
and special relativity share the same mathematics.  This aspect was
already noted in the literature for the case of the Wigner
rotation~\cite{kitano89}.  The concept of the Wigner rotation comes
from the kinematics of special relativity, in which two successive
non-collinear Lorentz boosts do not end up with a boost.  The result
is a boost followed or preceded by a rotation.  Thus we can achieve a
rotation from three non-collinear boosts starting from a particle at
rest.  Since each boost corresponds to an attenuation filter, it
requires three attenuation filters to achieve a Wigner rotation in
polarization optics.

While the Wigner rotation is based on Lorentz transformations of massive
particles, there are similar transformations for massless particles.
Here, two non-collinear Lorentz boosts do not result in one boost.
They become one boost preceded or followed by a transformation which
corresponds to a gauge transformation.  In two-by-two formalism, the
transformation takes the form of Eq.(\ref{shear1}).  We shall show in
this paper that the filter possessing the property of Eq.(\ref{shear1})
can be constructed from one rotation filter and one attenuation filter.
In mathematics, this type of decomposition is called the Iwasawa
decomposition~\cite{iwa49,simon98}.

While the primary purpose of this paper is to discuss filters and
their combinations in polarization optics, we provide also concrete
illustrative examples of Wigner's little group~\cite{wig39}.  The
little group is the maximal subgroup of the Lorentz group whose
transformations leave the four-momentum of a given particle invariant,
and has a long history~\cite{knp86}.  The Wigner rotation and the
Iwasawa decomposition are transformations of the little groups for
massive and massless particles respectively.  It is interesting to
note that these transformations can also be achieved in optics
laboratories.

In Sec.~\ref{formul}, we review the formalism for optical filters based
on the Lorentz group and explain why filters are like Lorentz
transformations.  It is shown in Sec. \ref{wigrot}, that a rotation can
be achieved by three non-collinear Lorentz boosts.  In Sec.~\ref{iwasa},
we spell out in detail how the Iwasawa decomposition can be achieved
from the combination of two optical filters.

\section{Formulation of the Problem}\label{formul}
In studying polarized light propagating
along the $z$ direction, the traditional approach is to consider the $x$
and $y$ components of the electric fields.  Their amplitude ratio and the
phase difference determine the degree of polarization.  Thus, we can
change the polarization either by adjusting the amplitudes, by changing
the relative phases, or both.  For convenience, we call the optical
device which changes amplitudes an ``attenuator'' and the device which
changes the relative phase a ``phase shifter.''

Let us write these electric fields as
\begin{equation}\label{expo1}
\pmatrix{E_{x} \cr E_{y}} =
\pmatrix{A \exp{\left\{i(kz - \omega t + \phi_{1})\right\}}  \cr
B \exp{\left\{i(kz - \omega t + \phi_{2})\right\}}} .
\end{equation}
where $A$ and $B$ are amplitudes which are real and positive
numbers, and $\phi_{1}$ and $\phi_{2}$ are the phases of the $x$ and
$y$ components respectively.  This column matrix is called the Jones
vector.  In dealing with light waves, we have to realize that the
intensity is the quantity we measure.  Then there arises the question of
coherence and time average.  We are thus led to consider the following
parameters.
\begin{eqnarray}\label{sii}
S_{11} &=& <E_{x}^{*}E_{x}>  , \qquad
S_{22} = <E_{y}^{*}E_{y}> , \cr
S_{12} &=& <E_{x}^{*}E_{y}> ,  \qquad
S_{21} = <E_{y}^{*}E_{x}> .
\end{eqnarray}
Then, we are naturally invited to write down the two-by-two matrix:
\begin{equation}\label{cohm1}
C = \pmatrix{<E^{*}_{x}E_{x}> & <E^{*}_{y} E_{x}> \cr
<E^{*}_{x} E_{y}> & <E^{*}_{y} E_{y}>} ,
\end{equation}
where $<E^{*}_{i}E_{j}>$ is the time average of $E^{*}_{i}E_{j}$.
The above form is called the coherency matrix~\cite{born80}.

It is sometimes more convenient to use the following combinations of
parameters.
\begin{eqnarray}\label{stokes}
&{}& S_{0} = S_{11} + S_{22}, \cr
&{}& S_{1} = S_{11} - S_{22}, \cr
&{}& S_{2} = S_{12} + S_{21}, \cr
&{}& S_{3} = -i\left(S_{12} - S_{21}\right).
\end{eqnarray}
These four parameters are called the Stokes parameters in the
literature~\cite{born80}.

We showed in our earlier papers that the Jones vectors and the
Stokes parameters can be formulated in terms of the two-by-two
spinor and four-by-four vector representations of the Lorentz group.
This group theoretical formalism allows to discuss three different
sets of physical quantities using one mathematical device.  In our
earlier publications, we used the concept of Lie groups extensively
and used their generators based on infinitesimal generators.

In this paper, we avoid the Lie groups and work only with explicit
transformation matrices.  For this purpose, we start with the
following two matrices.
\begin{eqnarray}
&{}& B = \pmatrix{\cosh\chi & \sinh\chi & 0 & 0 \cr
\sinh\chi & \cosh\chi & 0 & 0 \cr
0 & 0 & 1 & 0 \cr 0 & 0 & 0 & 1},  \nonumber \\[2ex]
&{}& R = \pmatrix{1 & 0 & 0 & 0 \cr 0 & \cos\phi & -\sin\phi & 0 \cr
0 & \sin\phi & \cos\phi & 0 \cr 0 & 0 & 0 & 1} .
\end{eqnarray}
If the above matrices are applied to the Minkowskian space of
$(ct, z, x, y)$, the matrix $B$ performs a Lorentz boost:
\begin{eqnarray}
&{}& t' = (\cosh\chi) t + (\sinh\chi) z , \nonumber \\[2ex]
&{}& z' = (\sinh\chi) t + (\cosh\chi) z,
\end{eqnarray}
while $R$ leads to a rotation:
\begin{eqnarray}
&{}& z' = (\cos\phi) z - (\sin\phi) x , \nonumber \\[2ex]
&{}& x' = (\sin\phi) z + (\cos\phi) x .
\end{eqnarray}
In our previous paper, we discussed in detail what these matrices do
when they are applied to the Stokes four-vectors.

In the two-component spinor space, the above transformation matrices
take the form
\begin{equation}
\pmatrix{e^{\chi/2} & 0 \cr 0 & e^{-\chi/2}}, \qquad
\pmatrix{\cos(\phi/2) & -\sin(\phi/2)
\cr \sin(\phi/2) & \cos(\phi/2)} .
\end{equation}
We discussed the effect of these matrices on the Jones spinors in our
earlier publications.

In this paper, we discuss some of nontrivial consequences
derivable from the algebra generated by these two sets of matrices.
We shall study Wigner rotations and Iwasawa decompositions.  The
Wigner rotation has been discussed in optical science in connection
with Berry's phase, but the Iwasawa decomposition is a relatively
new word in optics.  We would like to emphasize here that both
the Wigner rotation and Iwasawa decomposition come from the concept
of subgroup of the Lorentz groups whose transformations leave the
momentum of a given particle invariant.

\section{Wigner Rotations}\label{wigrot}
There are many different versions of the Wigner rotation in the literature.
Basically, this rotation is a product of two non-collinear Lorentz boosts.
The result of these two boosts is not a boost, but a boost preceded or
followed by a rotation.  This rotation is called the Wigner rotation.

\begin{figure}[thb]
\centerline{\includegraphics[scale=0.5]{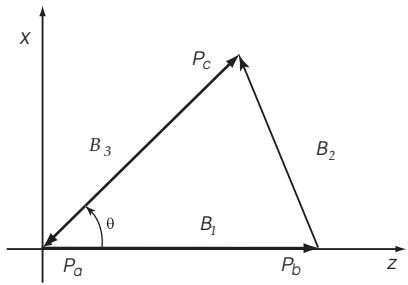}}
\caption{Closed Lorentz boosts.  Initially, a massive particle is at
rest with its four momentum $P_{a}$.  The first boost $B_{1}$ brings
$P_{a}$ to $P_{b}$.  The second boost $B_{2}$ transforms $P_{b}$ to
$P_{c}$.  The third boost $B_{3}$ brings $P_{c}$ back to $P_{a}$.
The particle is again at rest.  The net effect is a rotation around
the axis perpendicular to the plane containing these three
transformations.  We may assume for convenience that $P_{b}$ is along
the $z$ axis, and $P_{c}$ in the $zx$ plane.  The rotation is then
made around the $y$ axis.}
\end{figure}
In this paper, we approach the problem by using three boosts described
in Fig.~1.  Let us start with a particle at rest, with its four momentum
\begin{equation}\label{4a}
P_{a} = (m, 0, 0, 0),
\end{equation}
where we use the metric convention $(ct, z, x, y)$.  Let us next boost this
four-momentum along the $z$ direction using the matrix
\begin{equation}\label{boost1}
B_{1} = \pmatrix{\cosh\eta & \sinh\eta & 0 & 0 \cr
 \sinh\eta & \cosh\eta & 0& 0 \cr 0 & 0 & 1 &  \cr 0 & 0 & 0 & 1} ,
\end{equation}
resulting in the four-momentum
\begin{equation}\label{4b}
P_{b} = m (\cosh\eta, \sinh\eta, 0, 0) .
\end{equation}

Let us rotate this vector around the $y$ axis by an angle $\theta$.
Then the resulting four-momentum is
\begin{equation}\label{4c}
P_{c} = m \left(\cosh\eta, (\sinh\eta)\cos\theta,
(\sinh\eta)\sin\theta, 0 \right) .
\end{equation}
Instead of this rotation, we propose to obtain this four-vector by
boosting the four-momentum of Eq.(\ref{4b}).  The boost matrix in this
case is
\begin{eqnarray}\label{boost2}
&{}&  B_{2} = \pmatrix{1 & 0 & 0 & 0 \cr
0 & \cos\psi & -\sin\psi & 0 \cr
0 & \sin\psi & \cos\psi & 0 \cr 0 & 0 & 0 & 1}
\pmatrix{\cosh\lambda & \sinh\lambda & 0 & 0 \cr
\sinh\lambda & \cosh\lambda & 0 & 0 \cr
0 & 0 & 1 & 0 \cr 0 & 0 & 0 & 1} \nonumber \\[2ex]
&{}& \hspace{10mm} \times
\pmatrix{1 & 0 & 0 & 0 \cr 0 & \cos\psi & \sin\psi & 0 \cr
0 & -\sin\psi & \cos\psi & 0 \cr 0 & 0 & 0 & 1} ,
\end{eqnarray}
with
\begin{equation}\label{psi}
\lambda = 2 \tanh^{-1}\left\{[\sin(\theta/2)] \tanh\eta\right\} ,
\qquad \psi = {\theta \over 2} + {\pi \over 2} .
\end{equation}
If we carry out the matrix multiplication, the $B_{2}$ matrix becomes
\begin{equation}\label{boost22}
\pmatrix{\cosh\lambda & -\sin(\theta/2)\sinh\lambda  &
              \cos(\theta/2)\sinh\lambda  & 0\cr
-\sin(\theta/2)\sinh\lambda & 1 + \sin^{2}(\theta/2)(\cosh\lambda - 1)
              & -\sin\theta \sinh^{2}(\lambda/2) & 0 \cr
\cos(\theta/2) \sinh\lambda & -\sin\theta \sinh^{2}(\lambda/2) &
              1 + \cos^{2}(\theta/2)(\cosh\lambda - 1) & 0 \cr
0 & 0 & 0 & 1} .
\end{equation}
Next, we boost the four-momentum of Eq.(\ref{4c}) to that of Eq.(\ref{4a}).
The particle is again at rest.  The boost matrix is
\begin{eqnarray}\label{boost3}
&{}& B_{3} = \pmatrix{1 & 0 & 0 & 0 \cr
0 & \cos\theta & -\sin\theta & 0 \cr
0 & \sin\theta & \cos\theta & 0 \cr 0 & 0 & 0 & 1}
\pmatrix{\cosh\eta & -\sinh\eta & 0 & 0 \cr
-\sinh\eta & \cosh\eta & 0 & 0 \cr
0 & 0 & 1 & 0 \cr 0 & 0 & 0 & 1} \nonumber \\[2ex]
&{}& \hspace{10mm} \times
\pmatrix{1 & 0 & 0 & 0 \cr 0 & \cos\theta & \sin\theta & 0 \cr
0 & -\sin\theta & \cos\theta & 0 \cr 0 & 0 & 0 & 1} .
\end{eqnarray}
After the matrix multiplication,
\begin{equation}\label{boost33}
B_{3} = \pmatrix{\cosh\eta  & -\cos\theta \sinh\eta &
              -\sin\theta \sinh\eta & 0 \cr
-\cos\theta \sinh\eta & 1 + \cos^{2}\theta (\cosh\eta - 1) &
               \sin\theta \cos\theta (\cosh\eta - 1) & 0 \cr
-\sin\theta \sinh\eta & \sin\theta \cos\theta (\cosh\eta - 1) &
              1 + \sin^{2}\theta(\cosh\eta - 1) & 0 \cr
0 & 0 & 0 & 1} .
\end{equation}

The net result of these transformations is $B_{3}B_{2}B_{1}$.  This
leaves the initial four-momentum of Eq.(\ref{4a}) invariant.  Is it
going to be an identity matrix?  The answer is No.  The result of
the matrix multiplications is
\begin{equation}
W = \pmatrix{1 & 0 & 0 & 0 \cr 0 & \cos\Omega & -\sin\Omega & 0 \cr
   0 & \sin\Omega & \cos\Omega & 0 \cr 0 & 0 & 0 & 1} ,
\end{equation}
with
\begin{equation}\label{omeg}
\Omega = 2\, \sin^{-1}\left\{{(\sin\theta)\sinh^{2}(\eta/2) \over
\sqrt{\cosh^{2}\eta - \sinh^{2}\eta \sin^{2}(\theta/2)} }\right\} .
\end{equation}
This matrix performs a rotation around the $y$ axis and leaves the
four-momentum of Eq.(\ref{4a}) invariant.  This rotation is an
element of Wigner's little group whose transformations leave the
four-momentum invariant.  This is precisely the Wigner rotation.

This relativistic effect manifests itself in atomic spectra as the
Thomas precession.  Otherwise, the experiments on Wigner rotation
in special relativity is largely academic.  On the other hand, as
noted in the literature, this effect could be tested in optics
laboratories.  As for the Stokes parameters, the above four-by-four
matrices are directly applicable.   Indeed, each four-by-four matrix
corresponds to one optical filter applicable to polarized light.

In order to see this effect more clearly, let us use the Jones matrix
formalism.  The two-by-two  squeeze matrix corresponding to the boost
matrix $B_{1}$ of Eq.(\ref{boost1}) is
\begin{equation}
S_{1} = \pmatrix{e^{\eta/2} & 0 \cr 0 & e^{-\eta/2} } .
\end{equation}
The two-by-two squeeze matrix corresponding to the boost matrix of
Eq.(\ref{boost2}) is now
\begin{equation}
S_{2} = \pmatrix{\cos(\psi/2) & -\sin(\psi/2) \cr
          \sin(\psi/2) & \cos(\psi/2) }
        \pmatrix{e^{\lambda/2} & 0 \cr 0 & e^{-\lambda/2}}
        \pmatrix{\cos(\psi/2) & \sin(\psi/2) \cr
          -\sin(\psi/2) & \cos(\psi/2)} ,
\end{equation}
where the parameters $\psi$ and $\lambda$ are given in Eq.(\ref{psi}).
After the matrix multiplication, $S_{2}$ becomes
\begin{equation}
S_{2} = \pmatrix{\cosh(\lambda/2) - \sin(\theta/2) \sinh(\lambda/2) &
\cos(\theta/2) \sinh(\lambda/2) \cr
\cos(\theta/2) \sinh(\lambda/2) &
\cosh(\lambda/2) + \sin(\theta/2) \sinh(\lambda/2)} .
\end{equation}
This is a matrix which squeezes along the direction which makes
the angle $(\pi + \theta)/2$ with the $z$ axis.  The two-by-two
squeeze matrix corresponding to $B_{3}$ of Eq.(\ref{boost3}) is
\begin{equation}
S_{3} = \pmatrix{\cosh(\eta/2) - \cos\theta \sinh(\eta/2) &
         - \sin\theta \sinh(\eta/2) \cr
         - \sin\theta \sinh(\eta/2) &
          \cosh(\eta/2) + \cos\theta \sinh(\eta/2) } .
\end{equation}
Now the matrix multiplication $S_{3} S_{2} S_{1}$ corresponds to the
closure of the kinematical triangle given in Fig. 1.  The result is
\begin{equation}
S_{3}S_{2}S_{1} = \pmatrix{\cos(\Omega/2) & -\sin(\Omega/2) \cr
                  \sin(\Omega/2) & \cos(\Omega/2) } ,
\end{equation}
where $\Omega$ is given in Eq.(\ref{omeg}).

\section{Iwasawa Decompositions}\label{iwasa}
In Sec.~\ref{wigrot}, the Lorentz kinematics was based on a massive
particle at rest.  If the particle is massless, there are no Lorentz
frames in which the particle is at rest.  Thus, we start with a
massless particle whose momentum is in the $z$ direction:
\begin{equation}\label{ka}
K_{a} = (k, k, 0, 0) ,
\end{equation}
where $k$ is the magnitude of the momentum.   We can rotate this
four-vector to
\begin{equation}\label{kb}
K_{b} = (k, -k\sin\alpha, k\cos\alpha, 0)
\end{equation}
by applying to $K_{a}$ the rotation matrix
\begin{equation}\label{r+}
R_{+} = \pmatrix{1 & 0 & 0 & 0 \cr
0 & \cos\alpha_{+} & -\sin\alpha_{+}  & 0 \cr
0 & \sin\alpha_{+} & \cos\alpha_{+} & 0 \cr
0 & 0 & 0 & 1} ,
\end{equation}
with $\alpha_{+} = \alpha + \pi/2$.

\begin{figure}[thb]
\centerline{\includegraphics[scale=0.5]{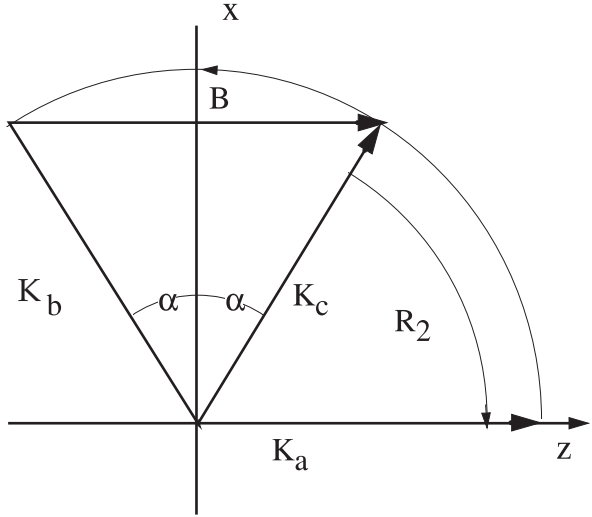}}
\caption{Two rotations and one Lorentz boost which preserve the
four-momentum of a massless particle invariant.  The four-momentum
$K_{a}$ is rotated to $K_{b}$ by $R_{+}$.  It is then boosted to
$K_{c}$ by the boost matrix $B$.  The rotation matrix $R_{-}$ brings
back the four-momentum to $K_{a}$.  The initial momentum is along the
$z$ direction, and the boost $B$ is also made along the same
direction.  The rotations are performed around the $y$ axis.}
\end{figure}

If we rotate $K_{b}$ around the $y$ axis by $-2\alpha$, the
resulting four-momentum will be
\begin{equation}\label{kc}
K_{c} = (k, k\sin\alpha, k\cos\alpha, 0)  .
\end{equation}
It is possible to transform $K_{b}$ to $K_{c}$ by applying to $K_{b}$
the boost matrix
\begin{equation}\label{boost4}
B = \pmatrix{\cosh\gamma & \sinh\gamma & 0 & 0 \cr
\cosh\gamma & \sinh\gamma & 0 & 0 \cr
0 & 0 & 1 & 0 \cr 0 & 0 & 0 & 1} .
\end{equation}
with
\begin{equation}\label{gamal}
\sinh\gamma = {2 \sin\alpha \over \cos^{2}\alpha} , \qquad
\cosh\gamma = {1 + \sin^{2}\alpha \over \cos^{2}\alpha } .
\end{equation}

We can transform $K_{c}$ to $K_{a}$ by rotating it around the $y$ axis
by $(\alpha - \pi/2)$.  The rotation matrix takes the form
\begin{equation}\label{r-}
R_{-} = \pmatrix{1 & 0 & 0 & 0 \cr
0 & \cos\alpha_{-}   & -\sin\alpha_{-}    & 0 \cr
0 & \sin\alpha_{-}  & \cos\alpha_{-}   & 0 \cr
0 & 0 & 0 & 1} ,
\end{equation}
with $\alpha_{-} = \alpha - \pi/2$.  Thus, the multiplication of
the three matrices, $R_{-} B R_{+}$, gives
\begin{equation}
T = \pmatrix{ 1 + u^{2}/2  & - u^{2}/2 & u  & 0    \cr
   u^{2}/2  & 1 - u^{2}/2 & u & 0   \cr
   u  & -u & 1 & 0 \cr
   0 & 0 & 0 & 1 } ,
\end{equation}
with
$$
u = - 2 \tan\alpha .
$$
This $T$ matrix plays an important role in studying space-time symmetries
of massless particles.  If this matrix is applied to the four-momentum
$K_{a}$ given in Eq. (\ref{ka}), the four-momentum remains invariant.
If this matrix is applied to the electromagnetic four-potential for
the plane wave propagating along the $z$ direction with the frequency
$k$, the result is a gauge transformation.

Again, the above four-by-four matrices are directly applicable to
the Stokes parameters.  On the other hand, if we are interested in
designing optical filters, we need two-by-two representations
corresponding to the four-by-four matrices given so far.  The
two-by-two squeeze matrix corresponding to the boost matrix $B$ of
Eq.(\ref{boost4}) is
\begin{equation}
S = \pmatrix{e^{\gamma/2} & 0 \cr 0 & e^{-\gamma/2}} ,
\end{equation}
while the two-by-two matrices corresponding to $R_{+}$ of Eq.(\ref{r+})
and $R_{-}$ of Eq.(\ref{r-}) are
\begin{equation}
R_{\pm} = \pmatrix{\cos(\alpha_{\pm}/2) & -\sin(\alpha_{\pm}/2) \cr
\sin(\alpha_{\pm}/2) & \cos(\alpha_{\pm}/2) } ,
\end{equation}
where $\alpha_{+}$ and $\alpha_{-}$ are given in Eq.(\ref{r+}) and
Eq.(\ref{r-}) respectively.  They satisfy the equations
$$
\alpha_{+} + \alpha_{-} = 2\alpha,  \qquad
\alpha_{+} - \alpha_{-} = \pi .
$$
The relation between $\gamma$ and $\alpha$ given in Eq.(\ref{gamal})
can also be written as $\cosh(\gamma/2) = 1/\cos\alpha$, which is
more useful for carrying out the two-by-two matrix algebra.

The matrix multiplication $R_{-} S R_{+} $ leads to
\begin{equation}\label{iwa1}
T = R_{-}SR_{+} = \pmatrix{1 & -2\,\tan\alpha \cr 0 & 1} .
\end{equation}
Conversely, we can write the
\begin{equation}
\pmatrix{1 & -2\,\tan\alpha \cr 0 & 1} = R_{-}SR_{+} .
\end{equation}
The $T$ matrix can be decomposed into rotation and squeeze matrices.
This possibility is called the Iwasawa decomposition.  In the present
case, $T$ of Eq.(\ref{iwa1}) can also be written as
\begin{equation}
T = R_{-}S \left\{\left(R_{-}\right)^{-1} R_{-1}\right\} R_{+}
= \left\{R_{-}S\left(R_{-}\right)^{-1}\right\}
       \left(R_{-} R_{+}\right) .
\end{equation}
The matrix chain $R_{-}S\left(R_{-}\right)^{-1}$ is one squeeze matrix
whose squeeze axis is rotated by $\alpha_{-}/2$, and the matrix product
$R_{-}R_{+}$ becomes one rotation matrix.  The result is
\begin{equation}
T = S(\alpha_{-}) R(2\alpha) ,
\end{equation}
with
\begin{eqnarray}
&{}& S(\alpha_{-}) =
       \pmatrix{\cosh(\gamma/2) + \cos\alpha_{-}\,\sinh(\gamma/2) &
       \sin\alpha_{-}\,\sinh(\gamma/2) \cr
       \sin\alpha_{-}\,\sinh(\gamma/2) &
  \cosh(\gamma/2) - \cos\alpha_{-}\,\sinh(\gamma/2) } , \nonumber \\[2ex]
&{}& R(2\alpha) = \pmatrix{\cos\alpha & -\sin\alpha \cr
\sin\alpha & \cos\alpha}.
\end{eqnarray}
It is indeed gratifying to note that the $T$ matrix can be decomposed
into one rotation and one squeeze matrix.  The squeeze is made along
the direction which makes an angle of $\alpha_{-}/2$ or
$-(\pi/2 - \alpha)/2$ with the $z$ axis.  The angle $\alpha$ is smaller
than $\pi/2$.

In our earlier papers~\cite{hkn97josa,hkn97pre}, we have discussed
optical filters with the property given in Eq.(\ref{iwa1}).  We said
there that filters with this property can be produced from an
infinite number of infinitely thin filters.  This argument was based
on the theory of Lie groups where transformations are generated by
infinitesimal generators.  This may be possible these days, but the
method presented in this paper is far more practical.  We need only
two filters~\cite{hks86jm}.

We are able to achieve this improvement because we used here the
analogy between polarization optics and Lorentz transformations which
share the same mathematical framework.

\section*{Concluding Remarks}
In this paper, we noted first that both the Wigner rotation and the
Iwasawa decomposition come from Wigner's little group whose
transformations leave the four-momentum of a given particle invariant.
Since the Lorentz group is applicable also to the Jones vector and
the Stokes parameters, it is possible to construct corresponding
transformations in polarization optics.  We have shown that both the
Wigner rotation and the Iwasawa decomposition can be realized in
optics laboratories.

The matrix of Eq.(\ref{shear1}) performs a shear transformation when
applied to a two-dimensional object, and has a long history in physics
and engineering.  It also has a history in mathematics.  The fact that
a shear can be decomposed into a squeeze and rotations is known
as the Iwasawa decomposition~\cite{iwa49}.

Among the many interesting applications of shear transformations, there
is a special class of squeezed states of photons or phonons having the
symmetry of shear~\cite{ky92,garr97}.  The wave-packet spread can be
formulated in terms of shear transformations~\cite{kiwi90aj}.

As we can see from this paper, a set of shear transformations can
be formulated as a subset of Lorentz transformations.  This set
plays an important role in understanding internal space-time
symmetry of massless particles, such as gauge transformation and
neutrino polarizations~\cite{knp86,wein64,hks82}.

\section*{Acknowledgments}
We would like to thank A. E. Bak for bringing to our attention to
his early works with C. S. Brown on applications of the Lorentz
group to polarization optics.  We are also grateful to S. Baskal
for telling us about the recent paper by Simon and Mukunda on
the Iwasawa decomposition~\cite{simon98}.

\end{document}